# Mouse Control using a Web Camera based on Colour Detection

**Abhik Banerjee[1], Abhirup Ghosh[2], Koustuvmoni Bharadwaj[3], Hemanta Saikia[4]**
*[1, 2, 3, 4]Department of Electronics & Communication Engineering,*
*Sikkim Manipal Institute of Technology*
*East Sikkim, India*

***ABSTRACT:*** *In this paper we present an approach for Human computer Interaction (HCI), where we have tried to control the mouse cursor movement and click events of the mouse using hand gestures. Hand gestures were acquired using a camera based on colour detection technique. This method mainly focuses on the use of a Web Camera to develop a virtual human computer interaction device in a cost effective manner.*

## I. INTRODUCTION

Human Computer Interaction today greatly emphasizes on developing more spontaneous and natural interfaces. The Graphical User Interface (GUI) on Personal Computers (PCs) is quiet developed, well defined and provides an efficient interface for a user to interact with the computer and access the various applications effortlessly with the help of mice, track pad, etc. In the present day scenario most of the mobile phones are using touch screen technology to interact with the user. But this technology is still not cheap to be used in desktops and laptops. Our objective was to create a virtual mouse system using Web camera to interact with the computer in a more user friendly manner that can be an alternative approach for the touch screen.

## II. RELATED WORKS

In reference [1], Erden et al have used a camera and computer vision technology, such as image segmentation and gesture recognition, to control mouse tasks. Our project was inspired by a paper of Hojoon Park [2] where he used Computer vision technology and Web camera to control mouse movements. However, he used finger-tips to control the mouse cursor and the angle between the thumb and index finger was used to perform clicking actions. Chu-Feng Lien [3] had used an intuitive method to detect hand motion by its Motion History Images (MHI). In this approach only finger tip was used to control both the cursor and mouse click. In his approach the user need to hold the mouse cursor on the desired spot for a specific period of time for clicking operation. Kamran Niyazi[4] et al used Web camera to detect color tapes for cursor movement. The clicking actions were performed by calculating the distance between two colored tapes in the fingers. In their paper [5] K N Shah et al have represented some of the innovative methods of the finger tracking used to interact with a computer system using computer vision. They have divided the approaches used in Human Computer Interaction (HCI) in two categories: 1. HCI without using interface and 2. HCI using interface. Moreover they have mentioned some useful applications using finger tracking through computer vision.

## III. INTRODUCTION TO THE SYSTEM

In our work, we have tried to control mouse cursor movement and click events using a camera based on colour detection technique. Here real time video has been captured using a Web-Camera. The user wears coloured tapes to provide information to the system. Individual frames of the video are separately processed. The processing techniques involve an image subtraction algorithm to detect colours. Once the colours are detected the system performs various operations to track the cursor and performs control actions, the details of which are provided below.

No additional hardware is required by the system other than the standard webcam which is provided in every laptop computer.

## IV. SYSTEM DESCRIPTION

Following are the steps in our approach:
(i) Capturing real time video using Web-Camera.
(ii) Processing the individual image frame.
(iii) Flipping of each image frame.
(iv) Conversion of each frame to a grey scale image.
(v) Colour detection and extraction of the different colours (RGB) from flipped gray scale image
(vi) Conversion of the detected image into a binary image.
(vii) Finding the region of the image and calculating its centroid.
(viii) Tracking the mouse pointer using the coordinates obtained from the centroid.
(ix) Simulating the left click and the right click events of the mouse by assigning different colour pointers.

4.1 The Basic Block Diagram of the System:

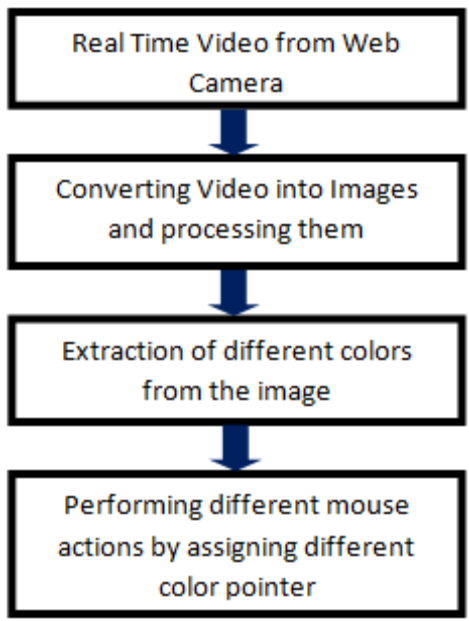

4.2 Capturing the real time video:

For the system to work we need a sensor to detect the hand movements of the user. The webcam of the computer is used as a sensor. The webcam captures the real time video at a fixed frame rate and resolution which is determined by the hardware of the camera. The frame rate and resolution can be changed in the system if required.

- Computer Webcam is used to capture the Real Time Video
- Video is divided into Image frames based on the FPS
  (Frames per second) of the camera
- Processing of individual Frames

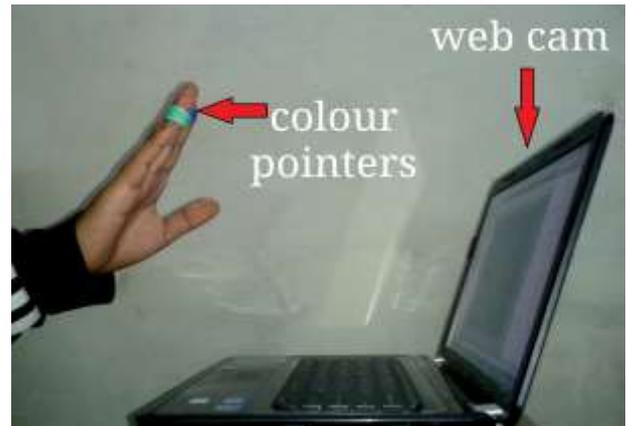

**Figure 1** Capturing the video

4.3 Flipping of Images:

When the camera captures an image, it is inverted. This means that if we move the colour pointer towards the left, the image of the pointer moves towards the right and vice-versa. It's similar to an image obtained when we stand in front of a mirror (Left is detected as right and right is detected as left). To avoid this problem we need to vertically flip the image. The image captured is an RGB image and flipping actions cannot be directly performed on it. So the individual colour channels of the image are separated and then they are flipped individually. After flipping the red, blue and green coloured channels individually, they are concatenated and a flipped RGB image is obtained. [6]

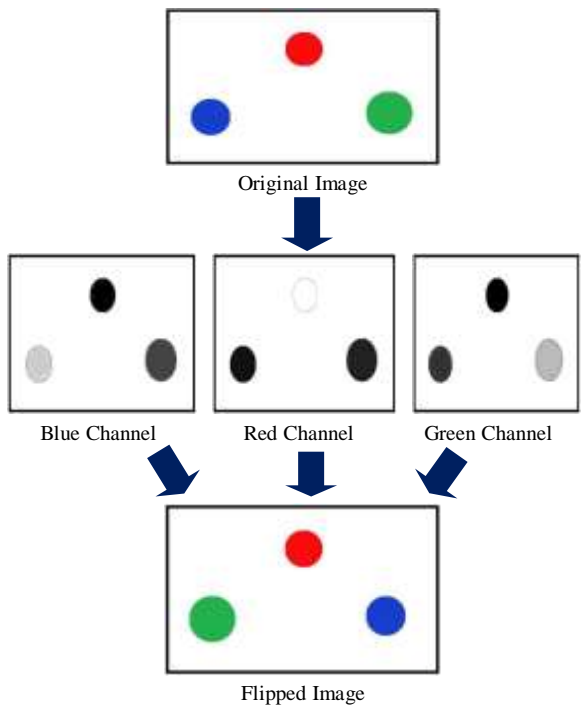

**Figure 2** Flipping of an Image

The following images show the entire flipping process in real time.

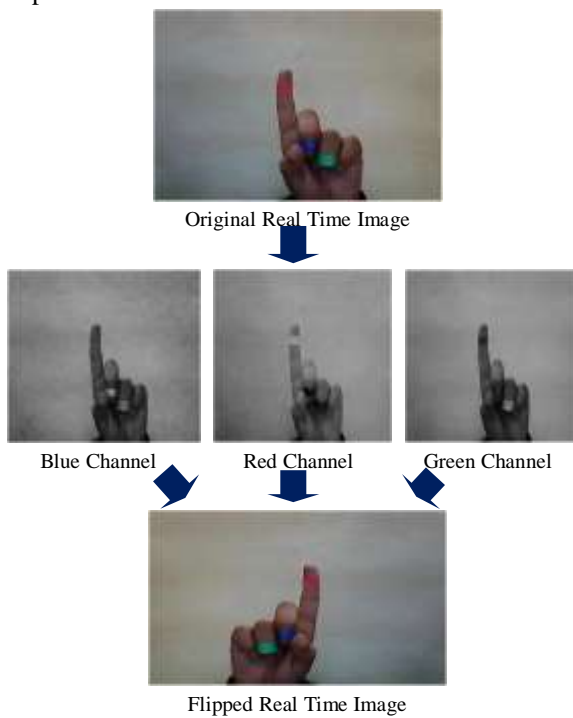

**Figure 3** Flipping of a Real time Image

4.4 Conversion of Flipped Image into Gray scale Image:

As compared to a coloured image, computational complexity is reduced in a gray scale image. Thus the flipped image is converted into a gray scale image. All the necessary operations were performed after converting the image into gray scale. [6]

4.5 Colour Detection:

This is the most important step in the whole process. The red, green and blue colour object is detected by subtracting the flipped color suppressed channel from the flipped Gray-Scale Image. This creates an image which contains the detected object as a patch of grey surrounded by black space. [7]

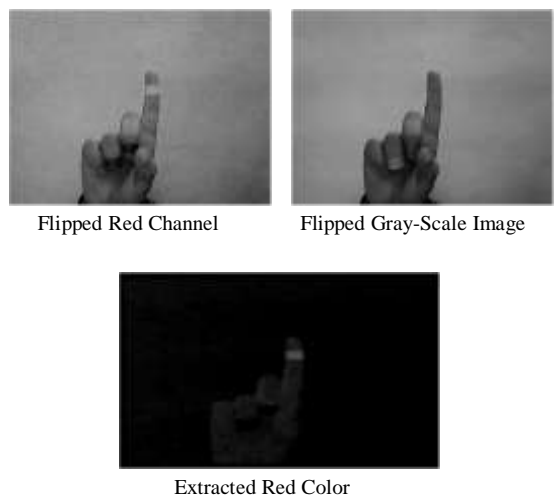

**Figure 4** Detection of Red Color

4.6 Conversion of gray scale Image into Binary scale Image:

The grey region of the image obtained after subtraction needs to be converted to a binary image for finding the region of the detected object. A grayscale image consists of a matrix containing the values of each pixel. The pixel values lay between the ranges 0 to 255 where 0 represents pure black and 255 represents pure white colour. We use a threshold value of 20% to convert the image to a binary image. This means that all the pixel values lying below 20% of the maximum pixel value is converted to pure black that is 0 and the rest is converted to white that is 1. Thus the resultant image obtained is a monochromatic image consisting of only black and white colours. The conversion to binary is required because MATLAB can only find the properties of a monochromatic image.

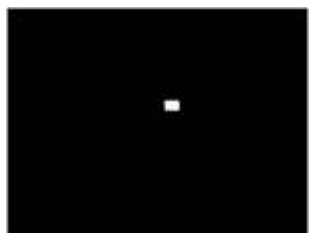

Detected Region

**Figure 5** The detected object which is associated with the mouse cursor

4.7    Finding Centroid of an object and plotting:

For the user to control the mouse pointer it is necessary to determine a point whose coordinates can be sent to the cursor. With these coordinates, the system can control the cursor movement.

An inbuilt function in MATLAB is used to find the centroid of the detected region. The output of function is a matrix consisting of the X (horizontal) and Y (vertical) coordinates of the centroid. These coordinates change with time as the object moves across the screen.
- Centroid of the image is detected
- Its co-ordinates are located and stored in a variable

4.8    Tracking the Mouse pointer:

Once the coordinates has been determined, the mouse driver is accessed and the coordinates are sent to the cursor. With these coordinates, the cursor places itself in the required position. It is assumed that the object moves continuously, each time a new centroid is determined and for each frame the cursor obtains a new position, thus creating an effect of tracking. So as the user moves his hands across the field of view of the camera, the mouse moves proportionally across the screen.

There is no inbuilt function in MATLAB which can directly access the mouse drivers of the computer. But MATLAB code supports integration with other languages like C, C++, and JAVA. Since java is a machine independent language so it is preferred over the others. A java object is created and it is linked with the mouse drivers. Based on the detection of other colours along with red the system performs the clicking events of the mouse. These colour codes can be customized based on the requirements. [8]

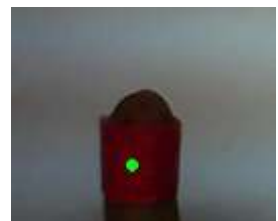

**Figure 6** Tracked Red Color

4.9    Performing Clicking Actions:

The control actions of the mouse are performed by controlling the flags associated with the mouse buttons. JAVA is used to access these flags. The user has to perform hand gestures in order to create the control actions. Due to the use of colour pointers, the computation time required is reduced. Furthermore the system becomes resistant to background noise and low illumination conditions. The detection of green and blue colours follows the same procedure discussed above.

Clicking action is based on simultaneous detection of two colours.
- If Red along with Green colour is detected, Left clicking action is performed
- If Red along with Blue colour is detected, Right clicking action is performed

**Mouse Control Flowchart:**

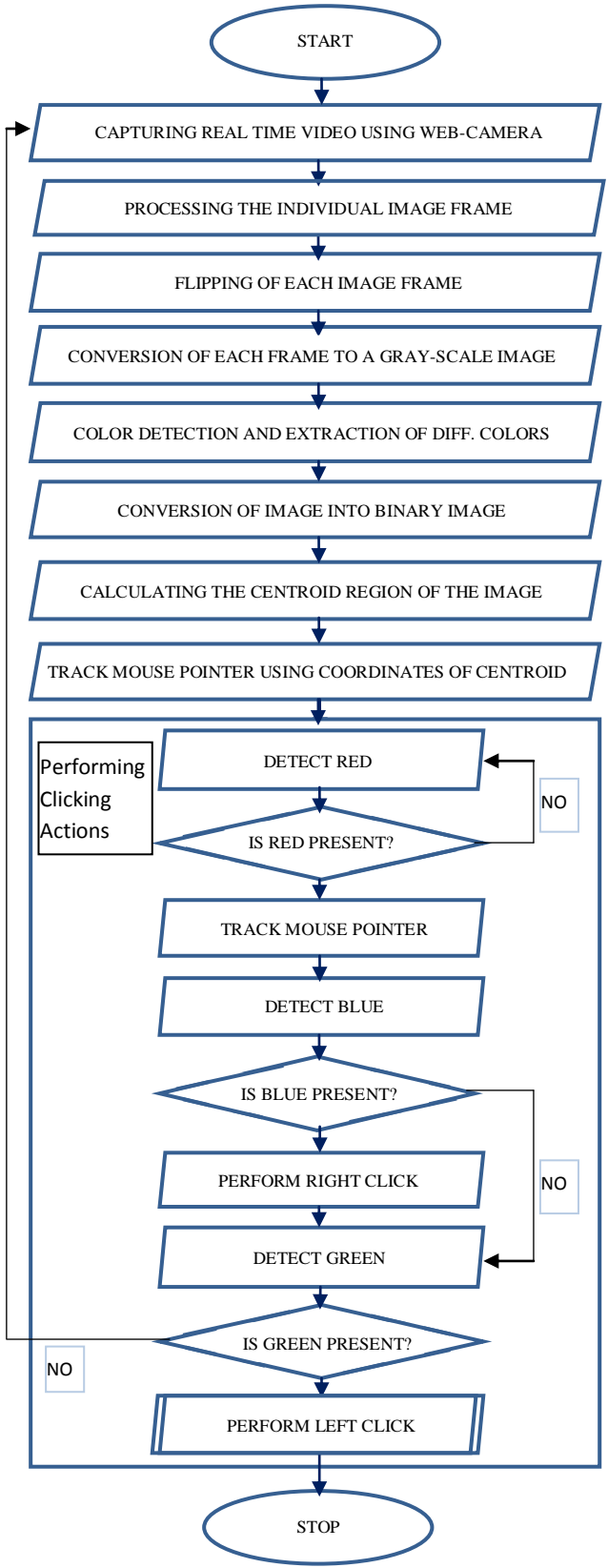

## V. PROBLEMS AND DRAWBACKS

Since the system is based on image capture through a webcam, it is dependent on illumination to a certain extent. Furthermore the presence of other colored objects in the background might cause the system to give an erroneous response. Although by configuring the threshold values and other parameters of the system this problem can be reduced but still it is advised that the operating background be light and no bright colored objects be present.

The system might run slower on certain computers with low computational capabilities because it involves a lot of complex calculations in a very small amount of time. However a standard pc or laptop has the required computational power for optimum performance of the system.

Another fact is that if the resolution of the camera is too high then the system might run slow. However this problem can be solved by reducing the resolution of the image by making changes in the system.

## VI. CONCLUSION

In this paper, an object tracking based virtual mouse application has been developed and implemented using a webcam. The system has been implemented in MATLAB environment using MATLAB Image Processing Toolbox.

This technology has wide applications in the fields of augmented reality, computer graphics, computer gaming, prosthetics, and biomedical instrumentation. Furthermore a similar technology can be applied to create applications like a digital canvas which is gaining popularity among artists. This technology can be used to help patients who don't have control of their limbs. In case of computer graphics and gaming this technology has been applied in modern gaming consoles to create interactive games where a person's motions are tracked and interpreted as commands.

Most of the applications require additional hardware which is often very costly. Our motive was to create this technology in the cheapest possible way and also to create it under a standardized operating system. Various application programs can be written exclusively for this technology to create a wide range of applications with the minimum requirement of resources.

## REFERENCES

[1] A. Erdem, E. Yardimci, Y. Atalay, V. Cetin, A. E. "Computer vision based mouse", Acoustics, Speech, and Signal Processing, Proceedings. (ICASS). *IEEE International Conference*, 2002

[2] Hojoon Park, "A Method for Controlling the Mouse Movement using a Real Time Camera", *Brown University*, Providence, RI, USA, Department of computer science, 2008

[3] Chu-Feng Lien, "Portable Vision-Based HCI – A Real-time Hand Mouse System on Handheld Devices", *National Taiwan University*, Computer Science and Information Engineering Department

[4] Kamran Niyazi, Vikram Kumar, Swapnil Mahe, Swapnil Vyawahare, "Mouse Simulation Using Two Coloured Tapes", Department of Computer Science, University of Pune, India, *International Journal of Information Sciences and Techniques (IJIST) Vol.2, No.2*, March 2012

[5] K N. Shah, K R. Rathod and S. J. Agravat, "A survey on Human Computer Interaction Mechanism Using Finger Tracking" *International Journal of Computer Trends and Technology, 7(3)*, 2014, 174-177

[6] Rafael C. Gonzalez and Richard E. Woods, *Digital Image Processing*, 2nd edition, Prentice Hall, Upper Saddle River, New Jersey, 07458

[7] Shahzad Malik, "Real-time Hand Tracking and Finger Tracking for Interaction", CSC2503F Project Report, December 18, 2003

[8] The MATLAB website. [Online]. Available: http://www.mathworks.com/matlabcentral/fileexchange/28757-tracking-red-color-objects-using-matlab